\documentclass[fleqn]{article}
\usepackage{makeidx}
\usepackage{psfig}
\usepackage{epsf}
\usepackage{graphicx}
\usepackage{isolatin1}

\usepackage{latexsym}
\usepackage{amssymb}
\usepackage{changebar}
\usepackage{booktabs}
\usepackage{amsmath}
\usepackage[small,bf]{caption}
\makeindex

\begin{document}

\newlength{\boxlength}
\setlength{\boxlength}{\textwidth}
\addtolength{\boxlength}{-2\parindent}

\newenvironment{algorithm}[1]
{
\begin{tabbing} xx \= xx \= xx \= xx \= xx \= xx \= xx \= xxxxx \kill
 {\bf algorithm} #1\\
 {\bf begin}\\
}
{
 {\bf end}\\
 \end{tabbing}
}

\title{Cluster Monte Carlo algorithms \thanks{ Chapter to appear in
`New Optimization Algorithms in Physics', 
edited by A. K. Hartmann and H. Rieger, 
(Wiley-VCh). ISBN: 3-527-40406-6, 
Estimated publishing date:    June 2004, 
Homepage: http://www.wiley-vch.de/publish/en/books/ }}
\author{Werner Krauth\\
CNRS-Laboratoire de Physique Statistique\\
Ecole Normale Sup\'{e}rieure\\
24, rue Lhomond\\
F-75231 Paris Cedex 05, France\\
Werner.Krauth@ens.fr}
\maketitle
\tableofcontents
\newcommand{\VEC}[1]{\mathbf{#1}}
\newcommand{\weq}[1]{eq.~(\ref{#1})}
\newpage

\section{Introduction}
In recent years, a better understanding of the Monte Carlo method has
provided us with many new techniques in different areas of statistical
physics.  Of particular interest are so called cluster methods, which
exploit the considerable algorithmic freedom given by the detailed
balance condition.  Cluster algorithms appear, among other systems,
in classical spin models, such as the Ising model \cite{wolff}, in
lattice quantum models (bosons, quantum spins and related systems)
\cite{evertz} and in hard spheres and other `entropic' systems for which
the configurational energy is either zero or infinite \cite{dresskrauth}.

In this chapter, we discuss the basic idea of cluster algorithms with
special emphasis on the pivot cluster method for hard spheres and related
systems, for which  several recent applications are presented.  We provide
less technical detail but more context than in the original papers.
The best implementations of the pivot cluster algorithm, the `pocket'
algorithm \cite{krauthmoessner}, can be programmed in a few lines.
We start with a short exposition of the detailed balance condition, and of
`a priori' probabilities, which are needed  to understand how optimized
Monte Carlo algorithms may be developed.  A more detailed discussion of
these subjects will appear in a forthcoming book \cite{smac}.

\section{Detailed balance and  a priori probabilities}

In contrast with the combinatorial optimization methods discussed
elsewhere in this book, the Monte Carlo approach does not construct a
well-defined state of the system ---minimizing the energy, or maximizing
flow, etc---but attempts to generate number of statistically independent
representative configurations $a$, with probability $\pi(a)$. In
classical equilibrium statistical physics, $\pi(a)$ is given by the
Boltzmann distribution, whereas, in quantum statistics, the weight is
the diagonal many-body density matrix.
\newcommand{\AAA}{\mathcal{A}}
\newcommand{\PP}{P}

In order to generate these configurations with the appropriate weight (and
optimal speed), the Monte Carlo algorithm moves (in one iteration) from
configuration $a$ to  configuration $b$ with probability $\PP(a\rightarrow
b)$.  This transition probability is chosen to satisfy the fundamental
condition of detailed balance
\begin{equation}
\pi(a) \PP(a \rightarrow b) = \pi(b) \PP(b \rightarrow a)
\label{e:detailed_balance}
\end{equation}
which is implemented using the Metropolis algorithm
\begin{equation}
\PP(a \rightarrow b) = \min \left(1, \frac{\pi(b)}{\pi(a)} \right)
\label{e:metropolis}
\end{equation}
or one of its variants.

\begin{figure}[h!t]
\begin{center}
\scalebox{1.0}{\includegraphics{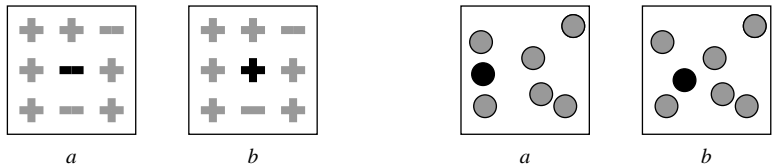}}
\caption{
Two examples of local Monte Carlo algorithms: the two-dimensional 
Ising model with single-spin flip dynamics (left) and two-dimensional
hard disks with a single-particle move (right).
}
\label{f:spin_model_hard_sphere}
\end{center}
\end{figure}

For the prototypical Ising model, the stationary probability distribution
(the statistical weight) of a configuration is the Boltzmann distribution
with an energy given by
\begin{equation}
E = - J \sum_{\langle i,j \rangle} S_i S_j\quad L > 0
\end{equation}
as used and modified in many other places in this book.  A common
move consists of a spin flip on a particular site $i$, transforming
configuration $a$ into another configuration $b$. This is shown
in Fig.~\ref{f:spin_model_hard_sphere} (left).  In a hard sphere
gas, also shown in Fig.~\ref{f:spin_model_hard_sphere} (right), one
typically displaces a single particle $i$ from $\VEC{x}$ to $\VEC{x}
+ \boldsymbol{\delta}$. There is a slight difference between these two
simple algorithms: by flipping the same spin twice, one goes back to the
initial configuration: a spin flip is its own inverse.  In contrast,
in the case of the hard-sphere system, displacing a particle twice by
the same vector $\boldsymbol{\delta}$ does not usually bring one back
to the original configuration.

An essential concept is the one of an a priori probability: it accounts
for the fact that the probability $\PP(a \rightarrow b)$ is a composite
object, constructed from the probability of \emph{considering} the move
from $a$ to $b$, and the probability of \emph{accepting} it.
\begin{equation}
\PP(a \rightarrow b) =
\underbrace{\AAA(a \rightarrow b)}_{\text{consider}\ a \rightarrow b }
\times
\underbrace{\tilde{\PP}(a \rightarrow b)}_{\text{accept}\ a \rightarrow b }
\nonumber
\end{equation}
In usual Monte Carlo terminology, if $a \rightarrow b$ is rejected
(after having been considered), then the `move' $a \rightarrow a$ is
chosen instead and the system remains where it is.

With these definitions, the detailed balance condition
\weq{e:detailed_balance} can be written as
\begin{equation}
\frac{\tilde{\PP}(a \rightarrow b)}{\tilde{\PP}(b \rightarrow a)} =
\frac{\pi(b)}{\AAA(a \rightarrow b)}
\frac{\AAA(b \rightarrow a)}{\pi(a)}
\nonumber
\end{equation}
and implemented by a Metropolis algorithm generalized from \weq{e:metropolis}:
\begin{equation}
\tilde{\PP}(a \rightarrow b) = 
\min \left\{1, \frac{\pi(b)}{\AAA(a \rightarrow b)}
\frac{\AAA(b \rightarrow a)}{\pi(a)} \right\}
\label{e:generalized_metropolis}
\end{equation}

It is very important to realize that the expression ``a priori probability
$\AAA(a \rightarrow b)$'' is synonymous to ``Monte Carlo algorithm''.
A Monte Carlo algorithm $ \AAA(a \rightarrow b)$ of our own conception
must satisfy three conditions:
\begin{enumerate}
\item It must lead the state of the system from a configuration $a$ to
a configuration $b$, in such a way that, eventually,  all configurations
in phase space can be reached (ergodicity).

\item It must allow to compute the ratio $\pi(a)/\pi(b)$. This is
trivially satisfied, at least for classical systems, as the statistical
weight is simply a function of the energy.

\item It must allow,  for any possible transition $a \rightarrow
b$, to compute both the probabilities $\AAA(a \rightarrow b)$ and
$\AAA(b\rightarrow a)$. Again, it is the ratio of probabilities which
is important.
\end{enumerate}

\begin{figure}[h!t]
\begin{center}
\scalebox{1.0}{\includegraphics{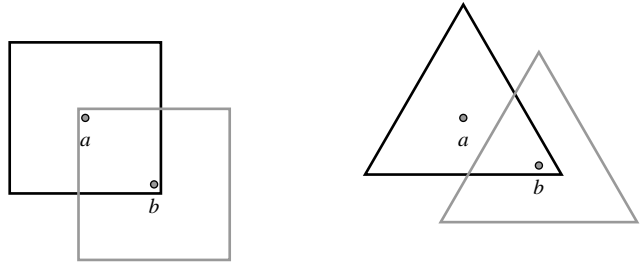}}
\caption{ A priori probabilities for the hard-sphere system. Left:
`square'---$\AAA(a\rightarrow b)$ is constant within the square
boundary, and zero outside.  By construction, $\AAA(a\rightarrow
b)=\AAA(b\rightarrow a)$. Right: `triangle'---for the analogous
(if hypothetical) case of a triangle, there are pairs $a, b$, where
$\AAA(a\rightarrow b)$ is finite, but $\AAA(b\rightarrow a)=0$. Both
rates $\PP(a\rightarrow b)$ and $\PP(b \rightarrow a)$ vanish.  }
\label{f:square_triangle}
\end{center}
\end{figure}

A trivial  application of a priori probabilities  for hard spheres
is given in Fig.~\ref{f:square_triangle}. (Suppose that the points
$a$ and $b$ are embedded in a large two-dimensional plane.)  On the
left side of the figure, we see one of the  standard choices for
the trial moves $\VEC{x} \rightarrow \VEC{x} + \boldsymbol{\delta}$
of a particle in Fig.~\ref{f:spin_model_hard_sphere}: The vector
$\boldsymbol{\delta}$ is uniformly sampled from a square centered around
the current position. If however, we decide, for some obscure reason,
to sample $\boldsymbol{\delta}$ from a triangle, we realize that in
cases such the one shown in Fig.~\ref{f:square_triangle} (right), the
a priori probability for the return move vanishes. It is easy to see
from \weq{e:generalized_metropolis} that, in this case, both $\PP(a
\rightarrow b)$ and $\PP(b \rightarrow a)$ are zero.

Notwithstanding its simplicity, the triangle `algorithm' illustrates
that \emph{any} Monte Carlo method $\AAA(a\rightarrow b)$ can
be made to comply with detailed balance, if we feed it through
\weq{e:generalized_metropolis}.  The usefulness of the algorithm is
uniquely determined by the speed with which it moves through configuration
space, and is highest if no rejections at all appear. It is to be noted
however that, even if $\tilde{\PP}(a\rightarrow b)$ is always $1$ (no
rejections), the simulation \emph{can} remain rather difficult. This
happens for example in the two-dimensional $XY$-model and in several
examples treated below.

Local algorithms are satisfactory for many problems but fail whenever
the typical differences between relevant configurations are much larger
than the change that can be achieved by one iteration of the Monte
Carlo algorithm.  In the Ising model at the critical point, for example,
the distribution of magnetizations is wide, but the local Monte Carlo
algorithm implements a change of magnetization of only $\pm 2$. This
mismatch lies at the core of critical slowing down in experimental
systems and on the  computer.

In liquids, modeled e.g. by the hard-sphere system, another well-known
limiting factor is that density fluctuations can relax only through
local diffusion.  This process generates slow hydrodynamic modes, if
the overall diffusion constants are small.

Besides these slow \emph{dense} systems, there is also the class of
highly \emph{constrained} models, of which binary mixtures will be treated
later. In these systems, the motion of some degrees of freedom naturally
couple to many others.  In a binary mixture, e.~g., a big colloidal
particle is surrounded by a large number of small particles, which are
influenced by its motion.  This is  extremely difficult to deal with
in Monte Carlo simulations, where the local moves $\VEC{x} \rightarrow
\VEC{x} + \boldsymbol{\delta}$ are essentially the unconstrained motion
of an isolated particle.

\section{The Wolff cluster algorithm for the Ising model}

The local spin-flip Monte Carlo algorithm not being satisfactory,
it would be much better to move large parts of the system, so called
clusters.  This cannot be done by a  blind flip of one or many spins
(with $\AAA(a\rightarrow b) = \AAA(b\rightarrow a)$), which allows
unit acceptance rate both for the move $a\rightarrow b$ and its reverse
$b \rightarrow a$ only if the energies of both configurations are the
same.  One needs an algorithm whose a priori probabilities ${\mathcal
A}(a \rightarrow b)$ and ${\mathcal A}(b \rightarrow a)$ soak up any
differences in statistical weight $\pi(a)$ and $\pi(b)$.

\begin{figure}[h!t]
\begin{center}
\scalebox{1.0}{\includegraphics{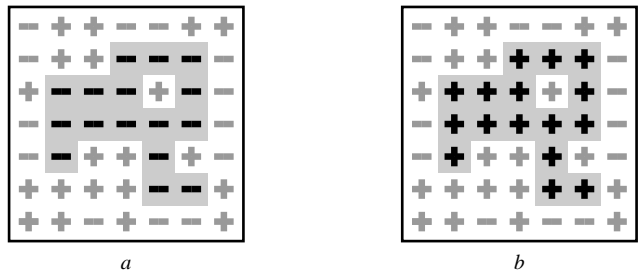}}
\caption{ The Wolff cluster algorithm for the Ising model adds, with
probability $p$, a link connecting a site outside the cluster to a site
already in the cluster (thereby adding the site). In the configuration
$a$, construction of the cluster (as shown) stopped with $9$ links
``$--$'', corresponding to an a priori probability $\AAA(a\rightarrow
b)=\AAA_{\text{interior}}\times (1-p)^9$. The return move stops
with probability $\AAA(b\rightarrow a)=\AAA_{\text{interior}}\times
(1-p)^{19}$, as there are $19$ links ``$++$''  across the boundary in
configuration $b$.  }
\label{f:spins_wolff}
\end{center}
\end{figure}

This can be done by starting the construction of a cluster with a
randomly sampled spin and by iteratively adding neighboring sites of the
same magnetization with a probability $p$.  To be precise, one should
speak about `links': if site $i$ is in the cluster and a neighboring
site $j$ is not, and if, furthermore, $S_i=S_j$, then one should add
link $\langle i,j\rangle$ with probability $p$. A site is added to the
cluster if it is connected by at least one link.  In configuration $a$
of Fig.~\ref{f:spins_wolff}, the cluster construction has stopped in the
presence of $9$ links ``$--$'' across the boundary. Each of these links
could have been accepted with probability $p$, but has been rejected.
This gives a term $(1-p)^9$ in the a priori probability.  Flipping the
cluster brings us to configuration $b$. The construction of the cluster
for configuration $b$ would  stop in the presence of $19$ links ``$++$''
across the boundary (a priori probability $\propto (1-p)^{19}$)).

This allows us to compute the a priori probabilities
\begin{align}
\AAA(a \rightarrow b) & =  \AAA_{\text{interior}}\times (1-p)^9& \notag\\
\AAA(b \rightarrow a) & =  \AAA_{\text{interior}}\times (1-p)^{19}& \notag\\
             E_a & =  E_{\text{interior}} + E_{\text{exterior}}  -9\times J + 19 \times J &
(\pi_a = \exp[-\beta E_a]) \notag\\
             E_b & =  E_{\text{interior}} + E_{\text{exterior}} - 19\times J +  9 \times J &
(\pi_b = \exp[-\beta E_b]) \notag
\end{align}
In these equations, the `interior' refers to the part of the cluster
which does not touch the boundary. By construction, the `interior' and
`exterior' energies and a priori probabilities are the same for any
pair of configurations $a$ and $b$ which are connected through a single
cluster flip.

We thus dispose of all the information needed to evaluate the acceptance
probability $\tilde{\PP}$ in \weq{e:generalized_metropolis}, which
we write more generally in terms of the number of ``same'' and of
``different'' links in the configuration $a$. These notions are
interchanged for configuration $b$  (in Fig.~\ref{f:spins_wolff}, we
have $ n_{\text{same}} = 9$, $n_{\text{diff}} = 19 $).  With the energy
scale $J$ set to $1$, we find
\begin{align} 
\tilde{\PP}(a\rightarrow b) & = \min\left\{1,          
\frac{e^{\beta n_{\text{diff}}} e^{-\beta n_{\text{same}}} }{(1-p)^{n_{\text{same}}}}
\frac{(1-p)^{n_{\text{diff}}}} {e^{-\beta n_{\text{diff}}} e^{\beta n_{\text{same}}} } 
\right\} \notag \\
& = \min \left\{1, 
\left[ \frac{e^{-2\beta}}{1-p} \right]^{n_{\text{same}}} 
\left[ \frac{1-p}{e^{-2\beta}} \right]^{n_{\text{diff}}} 
\right\}
\label{e:wolff_proba} 
\end{align}
Once the cluster construction stops, we know the configuration $b$,
may count $n_{\text{same}}$ and $n_{\text{diff}}$, and evaluate
$\tilde{\PP}(a\rightarrow b)$. Of course, a lucky coincidence 
\footnote{This accident explains the deep
connection between the Ising model and percolation.}  
occurs for $p=1-\exp[-2 J \beta]$.
This  special
choice yields a rejection-free algorithm whose acceptance probability
is unity for all possible moves and is implemented in the celebrated
Wolff cluster algorithm \cite{wolff}, the fastest currently known
simulation method for the Ising model. The Wolff algorithm can be
programmed in a few lines, by keeping a vector of cluster spins, and
an active frontier, as shown below.  The algorithm below presents a single 
iteration $a \rightarrow b$.  The function $\text{ran}[0,1]$ denotes
a uniformly distributed random number between $0$ and $1$, and $p$ is
set to the magical value $p=1- \exp [ -2 J \beta]$. The implementation
uses the fact that a cluster can grow only at its frontier (called the
`old' frontier $\mathcal{F}_{\text{old}}$, and generating the new one
$\mathcal{F}_{\text{new}}$).  It goes without saying that for the magic
value of $p$ we do not have to evaluate $\tilde{\PP}(a \rightarrow b)$ in
\weq{e:wolff_proba}, as it is always $1$.  Any proposed move is accepted.
\begin{algorithm}{wolff-cluster}
\> $i:=\text{random particle}$; \\
\> $\mathcal{C}:=\{ i \}$; \\
\> $\mathcal{F}_{\text{old}}:=\{i \};$\\
\>{\bf while} $\mathcal{F}_{\text{old}} \neq \{ \}$ {\bf do}\\
\>{\bf begin}\\
\>\> $\mathcal{F}_{\text{new}}:=\{ \};$\\
\>\>{\bf for} $\forall\ i\ \in\ \mathcal{F}_{\text{old}}$ {\bf do}\\
\>\>{\bf begin}\\
\>\>\>{\bf for} $\forall\ j\ \text{neighbor of $i$ with $S_i=S_j$},\ j
 \not \in \mathcal{C} $ {\bf do}\\
\>\>\>{\bf begin}\\
\>\>\>\>{\bf if} $\text{ran}[0,1] < p$ {\bf then}\\
\>\>\>\>{\bf begin}\\
\>\>\>\>\>$\mathcal{F}_{\text{new}}:= \mathcal{F}_{\text{new}}\cup \{j\};$\\
\>\>\>\>\>$\mathcal{C}:= \mathcal{C} \cup \{j\};$\\
\>\>\>\>{\bf end}\\
\>\>\>{\bf end}\\
\>\>{\bf end}\\
\>\>$\mathcal{F}_{\text{old}}:= \mathcal{F}_{\text{new}};$\\
\>{\bf end}\\
\>{\bf for} $\forall\ i\ \in\ \mathcal{C}$ {\bf do}\\
\>$S_i:= -S_i;$\\
\end{algorithm}

\section{Cluster algorithm for hard spheres and related systems}

We want to further exploit the analogy between the spin model and the 
hard-sphere system. As the spin-cluster algorithm constructs a cluster of spins
which \emph{flip} together, one might think that a cluster algorithm for
hard spheres should identify `blobs' of spheres that \emph{move} together.
Such a macroscopic ballistic motion would replace slow diffusion.

\begin{figure}[h!t]
\begin{center}
\scalebox{1.0}{\includegraphics{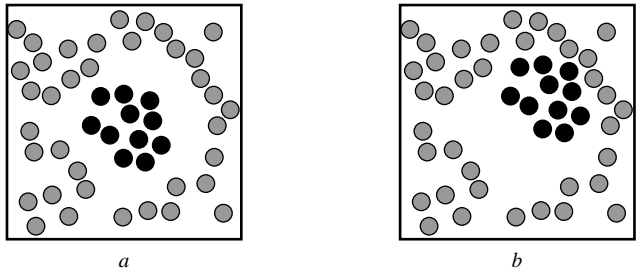}}
\caption{
The dark disks are easier to identify as a cluster in configuration
$a$ than in $b$, where they are fused into the background. This
means that, for the configurations $a$ and $b$ shown in this figure,
$\AAA(a \rightarrow b) \gg \AAA(b \rightarrow a)$
for any generic Monte Carlo algorithm. As $\pi(a)= \pi(b)$, the
acceptance probability $\tilde{\mathcal{\PP}}(a\rightarrow b)$ in
\weq{e:generalized_metropolis} will be extremely small. The problem
can be avoided \cite{dresskrauth} if the transformation $a\rightarrow b$ is protected by
a symmetry principle: it must be its own inverse.}
\label{f:hard_sphere_cluster}
\end{center}
\end{figure}

To see that this  strategy cannot be successful, it suffices to look
at the generalized detailed balance condition in the example shown in
Fig.~\ref{f:hard_sphere_cluster}: any reasonable algorithm $\AAA$ would
have less trouble spotting the cluster of dark disks in configuration $a$
than in $b$.  This means that $\AAA(a\rightarrow b) \gg \AAA(b \rightarrow
a)$ and that the acceptance rate $\tilde{\PP}(a \rightarrow b)$ would
be very small.

The imbalance between $\AAA(a\rightarrow b)$ and $\AAA(b\rightarrow a)$
can however be avoided if the two transition probabilities are protected
by a symmetry principle: the transformation $T$ producing $b$ from $a$
must be the same as the one producing $a$ from $b$.  Thus, $T$ should
be its own inverse.

\begin{figure}[h!t]
\begin{center}
\scalebox{1.0}{\includegraphics{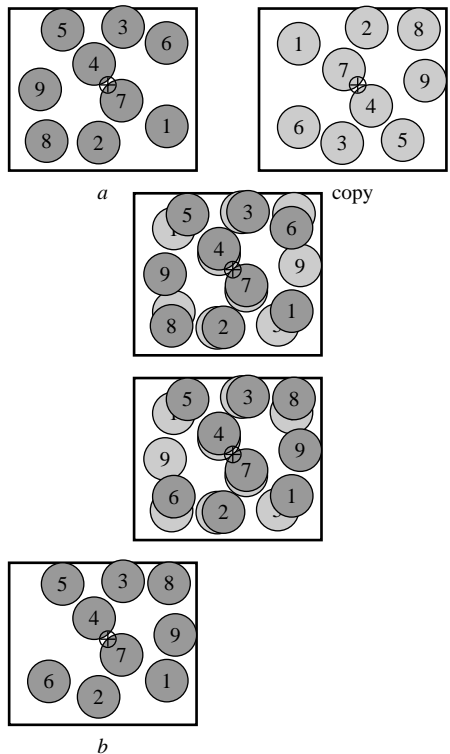}}
\caption{The pivot cluster algorithm performs a symmetry operation
which is its own inverse. In this system of hard disks (with periodic
boundary conditions), a rotation by an angle $\pi$ around an arbitrarily
sampled pivot ($\oplus$) is shown: $a$ is the original configuration, $b$
the rotated copy. The intermediate pictures show the superposed system of
original and copy before and after the flip.  The final configuration,
$b$, is also shown.  Notice that the transformation maps the simulation
box (with periodic boundary conditions) onto itself. If this is not
the case, the treatment of boundary conditions becomes more involved,
and generates rejections.}
\label{f:two_plates}
\end{center}
\end{figure}

In Fig.~\ref{f:two_plates}, this program is applied to a hard disk
configuration using as transformation $T$ a rotation by an angle $\pi$
around an arbitrarily sampled pivot (denoted by $\oplus$, for each
iteration a new pivot is used). Notice that for a symmetric particle,
the rotation by an angle $\pi$ is identical to the reflection around
the pivot.  It is useful to transform not just a single particle,
but the whole original configuration $a$ yielding the `copy'. By
overlaying the original with its rotated copy, we may identify the
invariant sub-ensembles (clusters) which transform independently under
$T$.  For example, in Fig.~\ref{f:two_plates}, we may rotate the disks
numbered $6$, $8$, and $9$, which form a cluster of overlapping disks
in the ensemble of overlayed original and copy.

In Fig.~\ref{f:two_plates}, there are the following three invariant
clusters:
\begin{equation}
\{6,8,9\}, \{2, 3, 4, 7 \}, \{1, 5\}
\label{e:invariant_sub_ensembles}
\end{equation}

The configuration $b$ in  Fig.~\ref{f:two_plates} shows the final
positions positions after rotation of  the first of these clusters. By
construction, $\AAA(a \rightarrow b)= \AAA(b \rightarrow a)$ and
$\pi(a)=\pi(b)$. This perfect symmetry ensures that detailed balance is
satisfied for the non-local move. Notice that moving the cluster $\{1,
5\}$ is equivalent to exchanging the labels of the two particles and
performing two local moves.  Ergodicity of the algorithm follows from
ergodicity of the local algorithm, as a local move $\VEC{x} \rightarrow
\VEC{x} + \boldsymbol{\delta}$ can always be disguised as a cluster
rotation around the pivot $\VEC{x} + \boldsymbol{\delta}/2$.

Fig.~\ref{f:two_plates} allows to understand the basic limitation of the
pivot cluster approach: if the density of particles becomes too large,
almost all particles will be in the same cluster, and flipping it will
essentially rotate the whole system. Nevertheless, even though above
the percolation threshold in the thermodynamic limit there exists
a large cluster containing a finite fraction of all particles, the
distribution of small clusters obeys an algebraic decay law. This means
that finite clusters of various sizes will be produced. These may give
rise to useful moves, for example in the case of dense polydisperse disks
discussed farther below. Even small clusters  provide non-diffusive mass
transport if they contain an odd number of particles (cf. the example
in Fig.~\ref{f:two_plates}) or particles of different type.

It is also useful to discuss what will happen if the ``copy''
does not stem from a symmetry operation, for example if the copy is
obtained from the original through a simple translation with a vector
$\boldsymbol{\delta}$.  In this case, there would still be clusters,
but they no longer appear in pairs. It would still be possible to flip
individual clusters, but not to conserve the number of particles  on
each plate.  This setting can also have important applications, it
is very closely related to Gibbs ensemble simulations and provides an
optimal way of exchanging particles between two plates.  The two plates
would no longer describe the same system but be part of a larger system
of coupled plates.

\begin{algorithm}{pocket-cluster}
\> $\VEC{r}_{\text{pivot}}:=\text{random point in box}$;\\
\> $i:=\text{random particle}$; \\
\> $\mathcal{P}:=\{ i \}$; \\
\> $\mathcal{O}:=\{\text{all particles} \} \setminus \{ i  \}$; \\
\>{\bf while} $\mathcal{P} \neq \{ \}$ {\bf do}\\
\>{\bf begin}\\
\>\> $i:=\text{any element of}\ \mathcal{P}$;\\
\>\> $\mathcal{P}:=\mathcal{P} \setminus \{i\}$;\\
\>\> $\VEC{r}(i):=\text{reflection of $\VEC{r}(i)$ around $\VEC{r}_{\text{pivot}}$}$;\\
\>\>{\bf for} $\forall\ j\ \in\ \mathcal{O}$ {\bf do}\\
\>\>{\bf if} $j\ \cap\ i$ {\bf then}\\
\>\>{\bf begin}\\
\>\>\> $\mathcal{O}:=\mathcal{O} \setminus \{ j  \}$;  \\
\>\>\> $\mathcal{P}:=\mathcal{P} \cup \{ j  \}$; \\
\>\>{\bf end}\\
\>{\bf end}\\
\end{algorithm}

Having discussed  the conceptual underpinnings of the pivot cluster
algorithm, it is interesting to understand how it can be made into
a working program. Fig.~\ref{f:two_plates} suggests one should use a
representation with two plates, and perform  cluster analyses, very
similar to what is done in the Wolff algorithm.

\begin{figure}[h!t]
\begin{center}
\scalebox{0.5}{\includegraphics{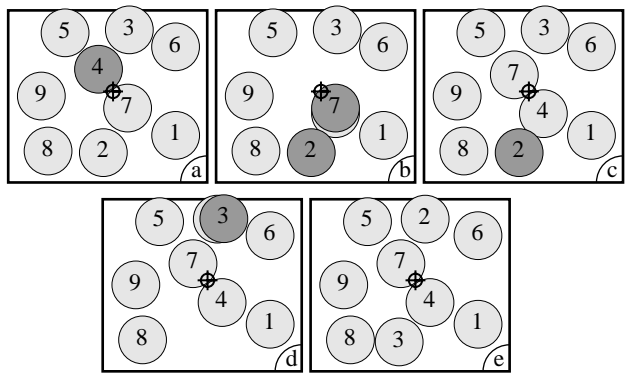}}
\caption{ One iteration of the pocket algorithm (`pocket' $\equiv$
`dark disks').  Initially  (frame $a$), a pivot is chosen and a starting
disk (here disk $4$) is put into the pocket. At each subsequent step,
a disk is removed from the pocket and transformed with respect to the
pivot. Any overlapping disks are added to the pocket. For example,
in frame $b$, overlaps exist between disk $4$ (which has just been
moved) and disks $2$ and $7$.  Only one of these disks is transformed in
frame $c$.  The pocket algorithm is guaranteed to move from valid
hard-disk configuration to another one, and to respect detailed balance.
It can be implemented in a few lines of code, as shown below.}
\label{f:pocket_disks}
\end{center}
\end{figure}

However, it is not necessary to work with two plates: The transformation
can be done on the system itself and does not even have to consider
a cluster at all. This ultimately simple solution is achieved in the
`pocket' algorithm \cite{krauthmoessner}: it merely keeps track of
particles which eventually have to be moved in order to satisfy all the
hard-core constraints: After sampling the pivot (or another symmetry
operation), one chooses a first particles, which is put into the pocket.
In each stage of the iteration, one particle is taken from the pocket, and
the transformation is applied to it. At the particle's new position, the
hard-core constraint will probably be violated for other particles. These
have simply to be marked as `belonging to the pocket'. One single  `move'
of the cluster algorithm  consists in all the stages until the pocket is
empty or, equivalently, in all the steps leading from frame $a$ to frame
$e$ in Fig.~\ref{f:pocket_disks}.  The inherent symmetry guarantees that
the process will end with an empty pocket, and detailed balance will
again be satisfied as the output is the same as in the two-plate version.

In the printed algorithm, $\mathcal{P}$ stands for the ``pocket'', and
$\mathcal{O}$ is the set of ``other'' particles that currently
do not have to be moved to satisfy the hard-core constraints.
The expression $j \cap i$ is `true' if the pair $i,j$ violates the
hard-core constraint.

\section{Applications}
\emph{Phase separation in binary mixtures}\\

\begin{figure}[h!t]
\begin{center}
\scalebox{1.0}{\includegraphics{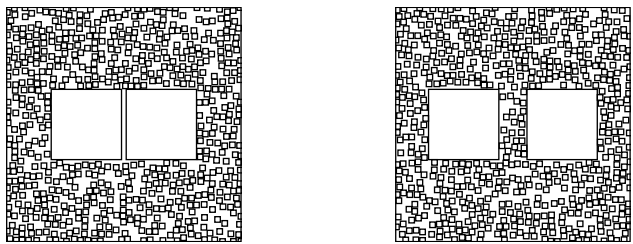}}
\caption{ Entropic interaction between two colloids (squares of
edge length $d_{\text{large}}$) in a sea of small particles (of size
$d_{\text{small}}$).  Left: Small particles cannot penetrate into the
slit between the large particles. The concentration difference leads to
an effective entropic interaction between colloids, which is attractive
at small separation.  Right: At large distances between colloids, the
effective interaction vanishes.  }
\label{f:pair_squares_close_wide}
\end{center}
\end{figure}
The depletion force---one of the basic interactions between colloidal
particles---is of purely entropic origin.  It is easily understood for
a system of large and small squares (or cubes): In the left picture of
Fig.~\ref{f:pair_squares_close_wide}, the two large squares are very
close together so that no small particles can penetrate into the slit
between the large ones. The finite concentration of small squares close
to the large squares constitutes a concentration (pressure) difference
between the outside and the inside, and generates an osmotic force  which
pulls the large squares together. The model of hard oriented squares
(or cubes) serves as an `Ising model of binary liquids' \cite{cuesta},
for which the force is very strong because of the large contact area
between them. Besides this, the situation is qualitatively similar to
the one for hard spheres.

For a long time, there was a dispute as to whether the depletion
interaction (which is oscillatory---repulsive and attractive, starting
with an attractive piece at small distances) was sufficiently strong
to induce phase transitions. The situation has been cleared up recently
due to experimental, analytical and numerical work.

We want to perform Monte Carlo simulation on this system
\cite{buhotkrauth1,buhotkrauth2}.  But as one can see immediately, this
is not simple: While the small squares may move with a local algorithm
of Fig.~\ref{f:spin_model_hard_sphere}, the large particles suffer from
a serious `pope in the crowd' effect: The large square is surrounded
by so many small particles in its immediate neighorhood that any trial
move will somewhere lead to the violation of the hard-core constraint,
i.e. it will be rejected.  A local Monte Carlo algorithm has  vanishing
acceptance rate for the motion of the large particles in the limit
of $d_{\text{small}}/d_{\text{large}} \rightarrow 0$, expressing the
increasing number of constraints in this limit.

\begin{figure}[h!t]
\begin{center}
\scalebox{0.92}{\includegraphics{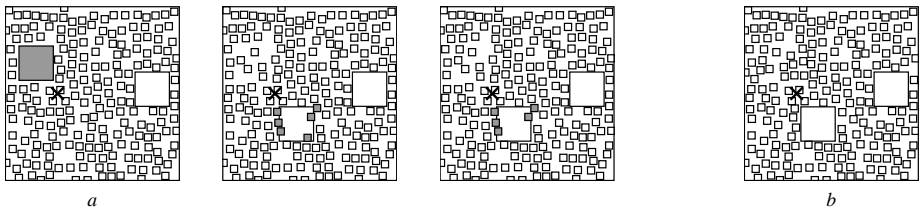}}
\caption{
Pocket algorithm applied to the binary mixture of squares.
The first three stages, and the final configuration of one step are shown.
Note that the squares which are fully covered by the moved large square
can be transformed immediately, without passing through the pocket, as they 
will not induce further overlaps.
}
\label{f:squares_cluster}
\end{center}
\end{figure}
The pivot  cluster method provides a straightforward solution to this
problem: randomly pick a square (large or small), and transform it by
applying a symmetry operation of the whole system (rotation around a
random pivot, reflection about a symmetry axis of the lattice).  At each
stage of the algorithm, pick an arbitrary particle from the pocket,
transform it and add to the pocket any particles it may overlap with.

As can be seen in Fig.~\ref{f:squares_cluster}, there is a nice
simplification: particles which are completely covered by a `big'
particle (as in the second frame of Fig.~\ref{f:squares_cluster})
will never generate new constraint violations. These particles can be
transformed directly, without passing through the pocket.

\begin{figure}[h!t]
\begin{center}
\scalebox{1.0}{\includegraphics{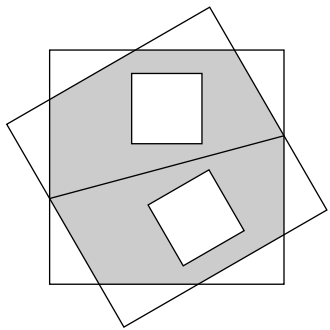}}
\caption{
The figure shows a transformation (reflection about a straight line)
which is not a symmetry transformation of the whole system (with periodic
boundary conditions). In this case, cluster transformations involving
squares outside the gray  area have to be rejected. Transformations as
the ones shown allow to reach arbitrary orientations of the squares.  }
\label{f:swivel}
\end{center}
\end{figure}
Using this algorithm, it has become possible to directly show that binary
mixtures undergo a phase separation transition, where the large particles
crystallize. The transition takes place at smaller and smaller densities
as the size mismatch $d_{\text{large}}/d_{\text{small}}$ increases at,
say, constant ratio of densities. At the same time, the percolation
threshold of the combined two-plate system is sensitive only to the
total density of particles.

It is also possible to relax the `orientation' constraint. This
can be done with transformations $T$  which satisfy  $T^2=1$, but
are not symmetries of the simulation box. An example is shown in
Fig.~\ref{f:swivel}.\\

\emph{Polydisperse mixtures}\\

\begin{figure}[h!t]
\begin{center}
\scalebox{1.0}{\includegraphics{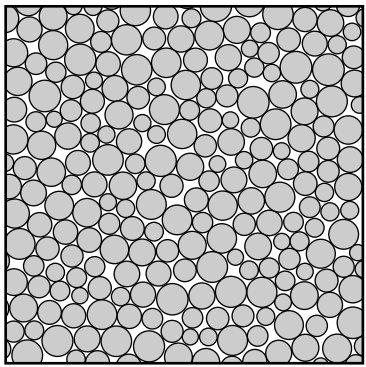}}
\caption{
Dense configuration of polydisperse hard disks, for which the time
evolution of a local Monte Carlo algorithm is immeasurably slow.  The
cluster algorithm remains ergodic at this density and even higher ones.
}
\label{f:dense_polydisperse}
\end{center}
\end{figure}

At several places in this chapter, the juxtaposition of spin systems
with hard spheres has lead to fruitful analogies.  One further analogy
concerns the very origin of the slowdown of the local algorithm.
In the Ising model, the critical slowing down is clearly rooted in the
thermodynamics of the system close to a second-order phase transition:
the distribution of the magnetization becomes wide, and the random walk
of the local Monte Carlo algorithm acquires a long auto-correlation time.

The situation is less clear, even extremely controversial, for the
case of hard-sphere systems. It is best discussed for polydisperse
mixtures, which avoid crystallization at high densities.  In
Fig.~\ref{f:dense_polydisperse}, a typical configuration of polydisperse
hard disks is shown at high density, where the time evolution of the
local Monte Carlo algorithm is already immeasurably slow. This system
behaves like a glass, and it is again of fundamental interest to study
whether there is a thermodynamic explanation for this, or whether the
system slows down for purely dynamic reasons.

In the spin problem, the cluster algorithms virtually eliminate
critical slowing down. These algorithms are the first to allow precision
measurements of thermodynamic properties close to the critical point.
The same has been found to apply for polydisperse hard disks, where the
pivot cluster algorithm and its variants allow perfect thermalization of
the system up to extremely high densities, even much higher than those
shown in Fig.~\ref{f:dense_polydisperse}.  As is evident from the figure,
the two-plate system is way beyond the percolation threshold, and one
iteration of the cluster algorithm likely involves a finite fraction
of all particles. The small clusters which are left behind lead to very
useful moves and exchanges of inequivalent particles.

Extensive simulations of this system have given no indications
of a thermodynamic transition.  For further discussion cf
\cite{santenkrauth1,santenkrauth2}. \\

\emph{Monomer-dimer problem}\\
\\
Monomer-dimer models are purely entropic lattice systems packed with hard
dimers (dominoes) which each cover two neighboring sites.  The geometric
cluster algorithm provides an extremely straightforward simulation method
for this system, for various lattices, and in two and higher dimensions
\cite{krauthmoessner}. In this case, the `clusters' have no branches. For
the completely covered dimer system (in the two-plate representation), the
clusters form closed loops, which are symmetric under the transformation.
These loops can be trivially generated with the pocket algorithm and
are special cases of transition graph loops used in other methods.

Care is needed to define the correct symmetry transformations. For
example, a pure rotation by an angle $\pi$ would leave the orientation
(horizontal, vertical) of each dimer unchanged, and conserve their
numbers separately.  On a square lattice of size $L\times L$, the
diagonals are symmetries of the whole system. It has been found that the
use of  reflections about all symmetry axes on the square or triangular
lattice leads to an ergodic algorithm. The reasoning can be extended to
higher dimensions\cite{huse}.  It is very interesting to observe that,
in any dimension, the cluster can touch the symmetry axis (or symmetry
hyperplane) at most twice.  This implies that symmetry axes (or their
higher dimensional generalizations) will not allow the cluster to fill
up the whole system. For a detailed discussion, cf. \cite{krauthmoessner}.

\begin{figure}[h!t]
\begin{center}
\scalebox{1.0}{\includegraphics{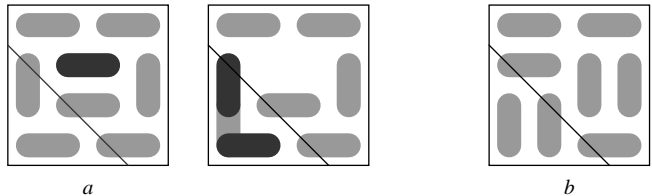}}
\caption{
Application of the pocket algorithm to a dimer-configuration on the 
two-dimensional square lattice. In this problem, the maximum size of the 
pocket is $2$. The initial configuration $a$, the configuration after the 
first transformation, and the final configuration $b$ are shown.  
}
\label{f:dimer_algo}
\end{center}
\end{figure}

\section{Limitations and Extensions}

As other powerful methods, the pivot cluster algorithm allows to solve
basic computational problems for some systems, but fails abysmally
for the vast majority. The main reason for failure is the presence of
clusters which are too large, in applications where they leave only
`uninteresting' small clusters.

This phenomenon is familiar from spin-cluster algorithms, which,  for
example fail for frustrated or random spin models, thus providing strong
motivation for many of the combinatorial techniques presented elsewhere
in this book. Clearly, a single method cannot be highly optimized and 
completely general at the same time.

In the first place, the cluster pivot algorithm has not improved
the notoriously difficult simulations for monodisperse hard disks at
the liquid-solid transition density. This density is higher than the
percolation threshold of the combined two-plate system comprising the
original and the copy. Nevertheless, one might suppose that the presence
of small clusters would generate fast non-local density fluctuations.
Unfortunately, this has not been found to have much impact on the overall
convergence times. A clear explanation of this finding is missing.

Another frustrating example is the Onsager problem of liquid crystals:
hard cylindrical rods with diameter $D$, and length $L$, which undergo
a isotropic-nematic transition at a volume fraction which goes to zero
as the rods become more and more elongated \cite{degennes}.
\begin{equation}
\rho_{\text{iso}} \sim 3.3 \frac{D}{L}\quad \text{for $D/L \rightarrow 0$}
\end{equation}
This is analogous to what we found for binary mixtures, where the
transition densities also go to zero with the ratio of the relevant
length scales,  and  one might think that the cluster algorithm should
work just as well as it does for binary mixtures.

\begin{figure}[h!t]
\begin{center}
\scalebox{1.0}{\includegraphics{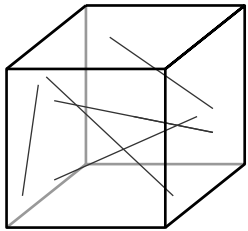}}
\caption{ Hard rods of length $L$ and diameter $D$ in a test box
of dimensions $L^3$. At the critical density for the isotropic-nematic
transition, the volume fraction occupied by the rods goes to zero, but the
cube is still opaque. This is due to the fact that the surface of a very
thin object ($\sim L D$) is much larger than its volume ($\sim L D^2$).  }
\label{f:hard_rods}
\end{center}
\end{figure}

Consider however a cube with edges of length $L$, filled with  density
$\rho_{\text{iso}}$ of rods (cf. Fig.~\ref{f:hard_rods}). The question of
the percolation threshold translates into asking what is the probability
of another, identical, rod to hit one of the rods in the system.
\begin{align}
\text{Volume of rods in cube of size}\  L^3  &\sim 3.3\  D L^2 \notag \\
\text{Number of rods }\ &\sim \frac{13.2}{\pi}\ L/D \notag \\
\text{Surface }\  &\sim \frac{13.2}{\pi}\  L^2 \notag
\end{align}

During the performance of the cluster algorithm, an external rod will be
moved into the test cube from elsewhere in the system. It is important
that it does not generate a large number $n_{\text{rods}}$ of violations
of the hard-core constraint with rods in the cube. We can orient the 
test cube such that the new rod comes in `straight' and find that the 
number is given as
\begin{equation}
n_{\text{rods}} \sim \frac{\text{surface of rods in test cube}}{\text{surface of test cube}}
\sim 4.2
\end{equation}
This is what was indeed observed: the exterior rod will hit around $4$
other rods, this means that this system is far above the percolation
threshold $n_{\text{rods}}=1$, and the cluster will contain essentially
all the rods in the system.

The pivot cluster algorithm has been used in a series of studies of more
realistic colloids, and has been extended to include a finite potential,
in addition to the hard-sphere interaction \cite{malherbe}.

Finally, the pivot cluster algorithm has  been very successfully applied
to the Ising model with fixed magnetization, where the number of ``$+$''
and of ``$-$'' spins are separately conserved. This is important in
the context of lattice gases, which can be mapped onto the  Ising model
\cite{heringa}.

\newcommand{\epl}{{Europhys.~Lett.}}
\newcommand{\jpa}{{J.~Phys.~A: Math~Gen.}}
\newcommand{\pra}{{Phys.~Rev.~A}}
\newcommand{\prb}{{Phys.~Rev.~B}}
\newcommand{\prc}{{Phys.~Rev.~C}}
\newcommand{\prd}{{Phys.~Rev.~D}}
\newcommand{\pre}{{Phys.~Rev.~E}}
\newcommand{\prl}{{Phys.~Rev.~Lett.}}

\section{Acknowledgments}
I would like to thank C. Dress, S. Bagnier, A. Buhot, L. Santen, and
R. Moessner for stimulating collaborations over the last
few years.

%\printindex

\begin{thebibliography}{99}

\bibitem{wolff}
U. Wolff, {\sl  Collective Monte Carlo Updating for Spin Systems},
\prl\ {\bf 62}, 361 (1989)

\bibitem{evertz}
H. G. Evertz, 
{\sl The Loop Algorithm},
Adv. Phys. {\bf 52}, 1 (2003)

\bibitem{dresskrauth}
C. Dress, W. Krauth, 
{\sl Cluster Algorithm for hard spheres and related systems}, 
\jpa\ {\bf 28}, L597 (1995)

\bibitem{krauthmoessner}
W. Krauth, R. Moessner, 
{\sl Pocket Monte Carlo algorithm for classical doped dimer models},
\prb\ {\bf 67},  064503 (2003)

\bibitem{smac}
W. Krauth,  \emph{Statistical Mechanics: Algorithms and Computations}, 
(Oxford University Press, 2004)

\bibitem{cuesta} J. A. Cuesta, {\sl Fluid Mixtures of Parallel Hard Cubes},
\prl\ {\bf 76},  3742 (1996)

\bibitem{buhotkrauth1}
A. Buhot, W. Krauth,
{\sl Numerical Solution of Hard-Core Mixtures},
\prl\ {\bf 80},  3787 (1998)

\bibitem{buhotkrauth2}
A. Buhot, W. Krauth,
{\sl Phase Separation in Two-Dimensional Additive Mixtures},
\pre\ {\bf 59},  2939 (1999)

\bibitem{santenkrauth1}
L. Santen, W. Krauth,
{\sl Absence of Thermodynamic Phase Transition in a Model Glass Former},
Nature {\bf 405}, 550 (2000)

\bibitem{santenkrauth2}
L. Santen, W. Krauth,
{\sl Liquid, Glass and Crystal in Two-dimensional Hard disks},
cond-mat/0107459

\bibitem{huse}
D. A. Huse, W. Krauth, R. Moessner, S. L. Sondhi,
{\sl Coulomb and Liquid Dimer Models in Three Dimensions},
\prl\ {\bf 91} 167004 (2003)

\bibitem{degennes}
P. G. de Gennes,
\emph{The Physics of Liquid Crystals},
(Oxford University Press, 1974)

\bibitem{malherbe} J. G. Malherbe, S. Amokrane,
{\sl Asymmetric mixture of hard particles with Yukawa attraction between unlike ones:
a cluster algorithm simulation study},
Mol. Phys. {\bf 97}, 677 (1999)

\bibitem{heringa}
J. R. Heringa and H. W. J. Bl\"{o}te,
{\sl The simple-cubic lattice gas with nearest-neighbour exclusion: Ising universality},
Physica {\bf 232A}, 369 (1996)

\end{thebibliography}
\end{document}